# Femtosecond x rays link melting of charge-density wave correlations and light-enhanced coherent transport in YBa$_2$Cu$_3$O$_{6.6}$

M. Först[1,2], A. Frano[3,4], S. Kaiser[1,2], R. Mankowsky[1,2], C.R. Hunt[1,2,5], J.J. Turner[6], G.L. Dakovski[6], M.P. Minitti[6], J. Robinson[6], T. Loew[3], M. Le Tacon[3], B. Keimer[3], J.P. Hill[7], A. Cavalleri[1,2,8], and S.S. Dhesi[9]

[1]*Max-Planck Institute for the Structure and Dynamics of Matter, Hamburg, Germany*
[2]*Center for Free Electron Laser Science, Hamburg, Germany*
[3]*Max-Planck Institute for Solid State Research, Stuttgart, Germany*
[4]*Helmholtz-Zentrum Berlin für Materialien und Energie, Berlin, Germany*
[5]*Department of Physics, University of Illinois at Urbana-Champaign, Urbana, IL*
[6]*Linac Coherent Light Source, Stanford Linear Accelerator Center (SLAC) National Accelerator Laboratory, Menlo Park, CA*
[7]*Condensed Matter Physics and Materials Science Department, Brookhaven National Laboratory, Upton, NY*
[8]*Department of Physics, Clarendon Laboratory, University of Oxford, UK*
[9]*Diamond Light Source, Chilton, Didcot, United Kingdom*

ABSTRACT

We use femtosecond resonant soft x-ray diffraction to measure the optically stimulated ultrafast changes of charge density wave correlations in underdoped $YBa_2Cu_3O_{6.6}$. We find that when coherent interlayer transport is enhanced by optical excitation of the apical oxygen distortions, at least 50% of the in-plane charge density wave order is melted. These results indicate that charge ordering and superconductivity may be competing up to the charge ordering transition temperature, with the latter becoming a hidden phase that is accessible only by nonlinear phonon excitation.

Hole doping of cuprates removes the antiferromagnetic order of the parent compound and promotes unconventional high-temperature superconductivity. A key ingredient in determining the critical temperature $T_C$ is the competition between superconducting phase coherence and charge or spin orders. A vivid demonstration of this interplay is the frustration of interlayer coupling by charge stripes around 1/8-doping in single-layer materials like $La_{2-x}Sr_xCuO_4$ and $La_{2-x}Ba_xCuO_4$ [1,2,3,4].

Most recently, it was shown that the "flattening" of the superconducting-to-normal-state phase boundary in $YBa_2Cu_3O_{6+x}$ near x=0.6 (~12.5% hole concentration) is coincident with the appearance of biaxial charge density wave (CDW) order [5,6,7,8,9]. Similar observations have been made in further high-$T_C$ cuprates, including $Bi_2Sr_{2-x}La_xCuO_{6+\delta}$ [10] and $HgBa_2CuO_{4+\delta}$ [11]. The interplay of charge order competing with superconductivity appears then to be a general phenomenon in the physics of these systems.

Although pressure [12,13] and magnetic fields [14] have long been used to affect this interplay at low temperatures, only recently it was shown that high-frequency optical pulses achieve a qualitatively similar effect over larger temperature ranges. For example, coherent interlayer coupling was induced in the low-temperature stripe-ordered phase of $La_{1.8-x}Eu_{0.2}Sr_xCuO_4$ [15] and $La_{2-x}Ba_xCuO_4$ [16], likely caused by the melting of the charge stripe order [17].

In $YBa_2Cu_3O_{6+x}$, optical excitation of apical oxygen distortions has been shown to cause an even more striking effect, enhancing coherent interlayer transport below $T_C$ and inducing a transient state above $T_C$ with important similarities to the equilibrium superconductor [18]. This effect was recently shown to involve redistribution of the tunneling strength from the intra-bilayer to the inter-bilayer regions of the unit cell [19] and a rearrangement of the lattice structure that could not be achieved at equilibrium [20].

Here, femtosecond resonant soft x-ray diffraction (RSXD) is combined with time-resolved THz spectroscopy to measure the response of the in-plane charge order in $YBa_2Cu_3O_{6.6}$. We wish to establish if the enhancement of coherent interlayer coupling involves a reduction of CDW order. We show that prompt reduction of the CDW resonant soft x-ray diffraction peak occurs as the material is transformed into the coherent state, providing a key microscopic ingredient for this class of phenomena.

Detwinned samples of $YBa_2Cu_3O_{6.6}$ were synthesized by the self-flux method. The equilibrium *c*-axis optical properties at *T*=20 K, below the superconducting transition temperature $T_C$ = 62 K, are reported in Figure 1 for frequencies between 0.5 and 2.5 THz. Quasi single-cycle THz-frequency pulses were generated by either optical rectification or by a photoconductive antenna and measured after reflection from the sample by electro-optic sampling. The reflected field was referenced to the same measurement made above $T_C$ and to literature data. The equilibrium reflectivity displays the Josephson plasma edge (Fig. 1(b)), a signature of supercurrent oscillations between capacitively coupled $CuO_2$ bilayers. As this is a longitudinal plasma excitation and involves a zero crossing of Re(ε(ω)), a peak in the loss function $-\text{Im } 1/\epsilon(\omega)$ is also observed, as displayed in Fig. 1(c).

Upon excitation with 300-femtosecond long pulses at 15 μm-wavelength, made resonant with the $B_{1u}$ infrared-active lattice distortion (670 cm$^{-1}$) sketched in Figure 1(a) [21], the same optical properties observed at equilibrium below $T_C$ appeared transiently in the normal state. The lower panels in Fig. 1(b) and (c) report a representative example of the photo-induced optical properties [17], measured for $T$ =100 K and ~1 ps after excitation at a fluence of 4-mJ/cm$^2$. Details about the time-resolved THz probe experiment, including data analysis are described in the Supplemental Material [22]. A reflectivity edge and a peak in the loss function are observed at $\omega_{JPR}$, underscoring transient interlayer (short-range) superconducting coherence. These effects can be induced only up to the temperature scale at which quasi-static charge order is observed ($T_{CO}$~160 K), suggesting a link between the two phenomena. This is clearly seen in Fig. 2(b), in which the strength of the static charge order, as revealed by π-polarized x-ray diffraction in resonance with the Cu $L_3$-edge (931.5 eV) at the in-plane wave vector $q_{\parallel}$ ~ 0.31 (see Fig. 2(a)), is plotted alongside the strength (volume fraction) of the light-induced coherent state [18].

Femtosecond resonant soft x-ray diffraction experiments were carried out at the Stanford LCLS x-ray free electron laser (FEL) under the same excitation conditions. The sample was mounted onto the same in-vacuum diffractometer used for the measurements of Fig. 2(a), cooled to immediately above the critical temperature $T_C$=62 K and excited by the same 15-μm wavelength pulses used for the THz probe experiments of Figure 1. The FEL photon energy was tuned to the Cu $L_3$-edge and cut to 0.5-eV bandwidth by a grating monochromator. The diffracted x rays were detected as a function of pump-probe time delay using an avalanche photodiode, enabling pulse-to-pulse normalization to the incident x-ray intensity.

Figure 3(a) shows the steady-state measurement of CDW diffraction at $q_{||} \sim 0.315$ at the free electron laser. Although the noise level has clearly increased in comparison to the synchrotron radiation measurements shown in Figure 2(a), the CDW related diffraction peak was detected at the same position in reciprocal space and with about the same amplitude and width above the fluorescence background (see inset). Figure 3(b) shows the transient change of the peak amplitude, normalized to the steady state after subtraction of the fluorescence background. Here, we assume that the fluorescence background is not altered on the ultrafast time scale. After excitation, the scattering signal reduced promptly to approximately half of its equilibrium value. Because the x-ray pulses were absorbed over a 200-nm layer and the excitation pulse was deposited over a ~2-μm depth, the reduction in the scattering signal can be directly related to the melting of approximately 50 % of the charge order.

The observed disappearance of charge order occurs on a timescale comparable with the appearance of the plasma edge, strongly indicating a correlation between the two phenomena. However, whilst the transient plasma edge survives only for 5-7 ps [18], the charge order remains melted for a significantly longer time. This is most likely due to the fact that interlayer coherence disappears immediately after the local lattice distortions are relaxed [20], but the recovery of charge order requires the buildup of correlations on longer length scales.

The reported melting of charge order is reminiscent of the physical origin of light-induced interlayer coherence as for single-layer stripe ordered cuprates at low temperatures [15,17]. In the striped compounds the effect was easily understood by considering frustrated Josephson coupling due to pair density wave order [23], yet the present results in $YBa_2Cu_3O_{6.6}$ does not lend itself to an equally simple interpretation.

The data are strongly indicative of a ground state in which charge order, or a fluctuating/intertwined state involving both charge and superconducting order [24], frustrate superconductivity at least immediately above $T_C$. The microscopic physics following the optical excitation is likely to involve anharmonic lattice motions driven by the optical excitation [20, 25], or alternatively a more complex stabilization effect for the light-induced coherent tunneling in the high-temperature state. Light-induced coherence appears at higher temperatures than residual phase coherence in the planes [26], indicating the presence of a hidden state invisible in equilibrium. Similarly, charge order melting may also be invoked for the light-induced coherence obtained at lower doping values reported in Refs. 18 and 19, although in that case the ordering of charges appears on shorter range, and is not accessible with femtosecond x rays.

Theoretical efforts will be necessary to explain the observed experimental features in more detail. Future experimental work will focus on improvements in our ability to controlling light-induced melting of charge order, perhaps even minimizing dissipation to achieve steady state coherence by continuous wave light excitation.

## ACKNOWLEDGEMENT

Portions of this research were carried out on the SXR Instrument at the Linac Coherent Light Source (LCLS), a division of SLAC National Accelerator Laboratory and an Office of Science user facility operated by Stanford University for the U.S. Department of Energy. The SXR Instrument is funded by a consortium whose membership includes the LCLS, Stanford University through the Stanford Institute for Materials Energy Sciences (SIMES), Lawrence Berkeley National Laboratory (LBNL, contract No. DE-AC02-05CH11231), University of Hamburg through the BMBF priority program FSP 301, and the Center for Free Electron Laser Science (CFEL).

The research leading to these results has received funding from the European Research Council under the European Union's Seventh Framework Programme (FP7/2007-2013) / ERC Grant Agreement n° 319286 (Q-MAC). Work at Brookhaven National Laboratory was funded by the Department of Energy, Division of Materials Science and Engineering under contract No. DE-AC02-98CH10886.

**FIGURE CAPTIONS**

**Figure 1.** (a) Crystal structure of orthorhombic YBa$_2$Cu$_3$O$_{6.6}$ and the motion of the apical oxygen atoms (red shadows) associated with the resonantly excited *c*-axis B$_{1u}$ phonon mode. (b) Top: Below-$T_C$ (20 K) static frequency-dependent reflectivity of THz light polarized along the *c*-axis, clearly showing the Josephson plasma edge. The lower panel shows the light-induced reflectivity changes $\Delta R/R_0$ above $T_C$ (100 K, green dots). Here, the sample was excited with 300-fs pulses at 15 µm wavelength, polarized along the *c*-axis, and the data are taken at +0.8 ps time delay. At negative time delay (light green solid line), the sample does not react to the mid-infrared excitation. (c) The static above-$T_C$ electron loss function is shown at the top. The lower panel depicts the light-induced change in the *T>T$_C$* loss function for the same conditions as described in part (b).

**Figure 2.** (a) RSXD scan of the YBa$_2$Cu$_3$O$_{6.6}$ CDW peak at 62 K, using π-polarized x-rays at the Cu L$_3$-edge (blue data points), with $q_{//}$ the in-plane component of the diffraction wave vector along the (1 0 0) direction. The blue solid line is a polynomial fit to the fluorescence background. The inset shows the same diffraction peak normalized to this background and fitted with a Gaussian function. Data were taken using synchrotron radiation at the Diamond Light Source. (c) Temperature dependence of the integrated intensity of the CDW peak (blue circles) and of the volume fraction of the transient superconducting state.

**Figure 3:** (a) The same RSXD scan as shown in Fig. 2(a), now measured at the Linac Coherent Light Source free electron laser. (b) Transient height of the CDW diffraction peak induced by direct excitation of the apical oxygen mode using 400-fs pulses at 15 μm wavelength, polarized along the *c*-axis.

# FIGURES

# Figure 1

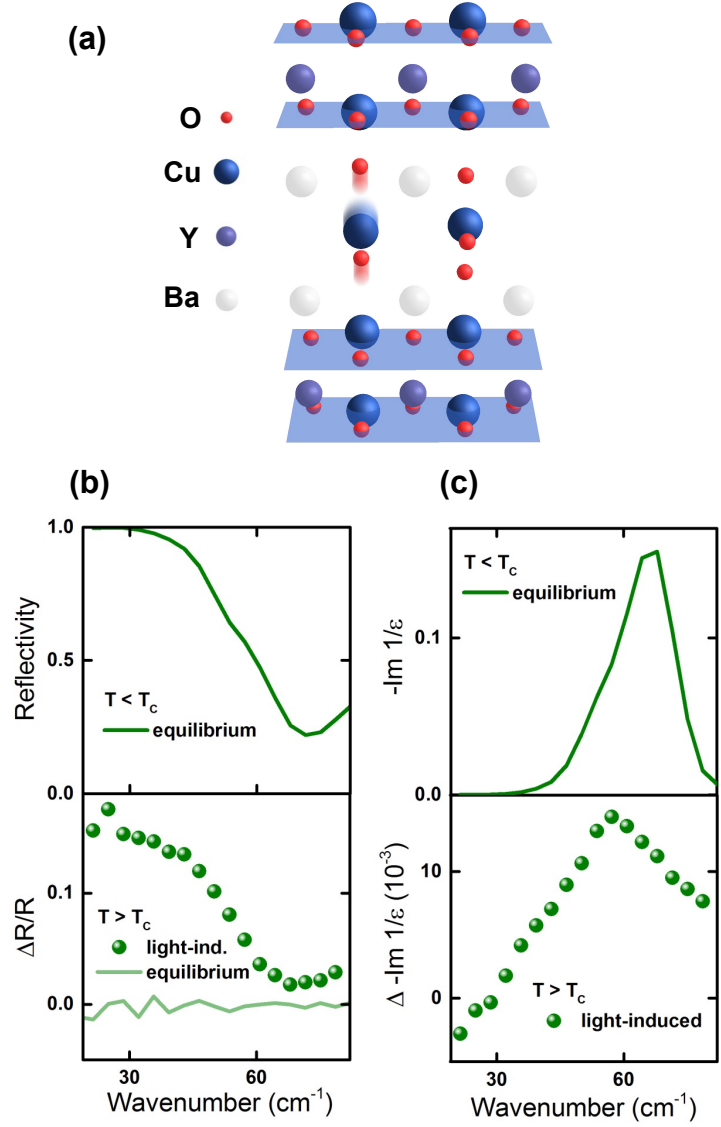

**Figure 2**

**(a)**

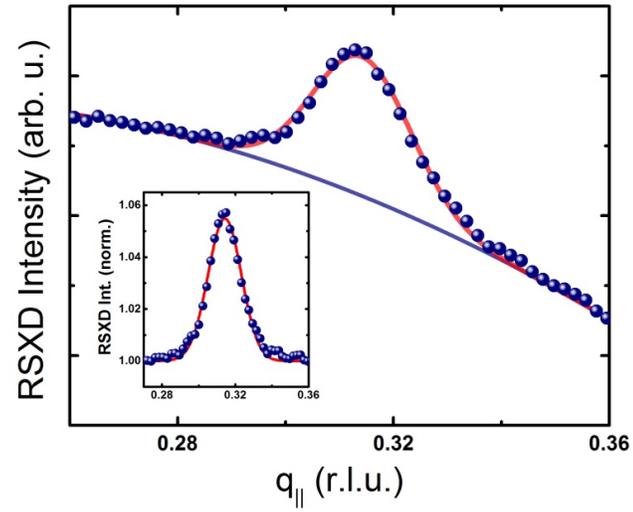

**(b)**

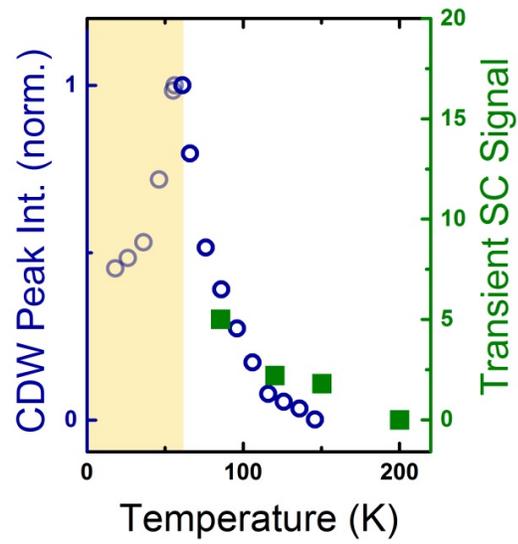

**Figure 3**

**(a)**
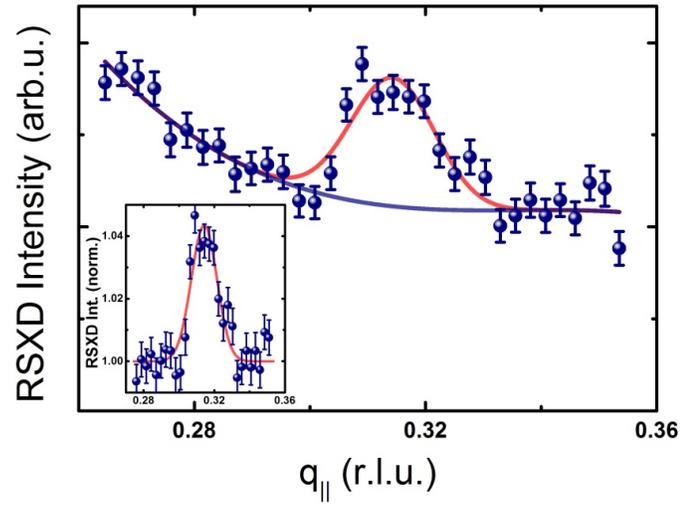

**(b)**
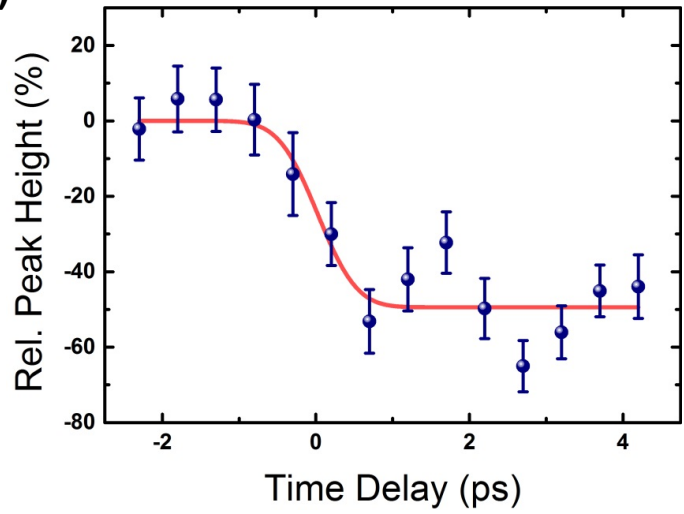

# Supplemental Information:

# Femtosecond x-rays link melting of charge density wave correlations and light-enhanced coherent transport in YBa$_2$Cu$_3$O$_{6.6}$

M. Först[1,2], A. Frano[3,4], S. Kaiser[1,2], R. Mankowsky[1,2], C.R. Hunt[1,2,5], J.J. Turner[6], G.L. Dakovski[6], M.P. Minitti[6], J. Robinson[6], T. Loew[3], M. Le Tacon[3], B. Keimer[3], J.P. Hill[7], A. Cavalleri[1,2,8], and S.S. Dhesi[9]

[1]Max-Planck Institute for the Structure and Dynamics of Matter, Hamburg, Germany
[2]Center for Free Electron Laser Science, Hamburg, Germany
[3]Max-Planck Institute for Solid State Research, Stuttgart, Germany
[4]Helmholtz-Zentrum Berlin für Materialien und Energie, Berlin, Germany
[5]Department of Physics, University of Illinois at Urbana-Champaign, Urbana, IL
[6]Linac Coherent Light Source, Stanford Linear Accelerator Center (SLAC) National Accelerator Laboratory, Menlo Park, CA
[7]Condensed Matter Physics and Materials Science Department, Brookhaven National Laboratory, Upton, NY
[8]Department of Physics, Clarendon Laboratory, University of Oxford, UK
[9]Diamond Light Source, Chilton, Didcot, United Kingdom

**S1: THz probe experiments**

The changes in the 0.5–2.5 THz optical conductivity perpendicular to the CuO$_2$ planes, induced by the mid-infrared resonant lattice excitation, were measured in the setup described in full detail in Ref. 1. In brief, near-infrared 100-fs pulses at 800 nm wavelength are split into three beams (i) to feed an optical parametric amplifier generating the mid-IR pump pulses, (ii) to generate the THz probe field via optical rectification in a ZnTe crystal, and (iii) to detect this THz field in amplitude and phase after reflection from the YBa$_2$Cu$_3$O$_{6.6}$ sample via electro-optic sampling in a second ZnTe crystal. For a given pump-probe time delay $\tau$ the relative delay between the mid-IR excitation and the 800-nm sampling pulse was kept fixed, and the THz transient was scanned with respect to these two. Therefore, each point in the THz profile probed the material at the same time delay $\tau$ after excitation.

The transient frequency-dependent optical properties of the photo-excited volume were then calculated by considering the equilibrium THz optical properties and the several-µm different penetration depths between the (mid-IR) pump and (THz) probe light. Here, it was considered that the pump-induced change in the optical properties is maximum at the sample surface and decays exponentially into the bulk according to Beer's law. The extinction depth for 15-µm pump light in resonance with the B$_{1u}$ phonon mode is about 1-2 µm. To evaluate the depth dependent changes in the optical properties of the far deeper (about 20 µm) penetrating THz probe pulses, a matrix tranfer formalism was used splitting the probing depth $L$ into a stack of layers of equal thickness $d$ much smaller than $L$.

Figure S1 shows the temperature dependence of the fully processed changes in the THz optical properties, following the mid-infrared excitation described in the main text. In addition, we are plotting a quantitative fit to these data (black solid lines) considering an inhomogeneously transformed pump volume consisting of a mixture of a high-mobility conductor (volume fraction $f$), with the dielectric function that contains the photo-induced plasma edge, and of unperturbed YBCO with the experimentally determined equilibrium optical properties of the normal state. The data show that with increasing temperature the system is less effectively driven into the state of coherent transport. The strength of the transient superconducting state, shown in Fig. 1(e) of the main paper, is deduced from this temperature dependent filling fraction $f$.

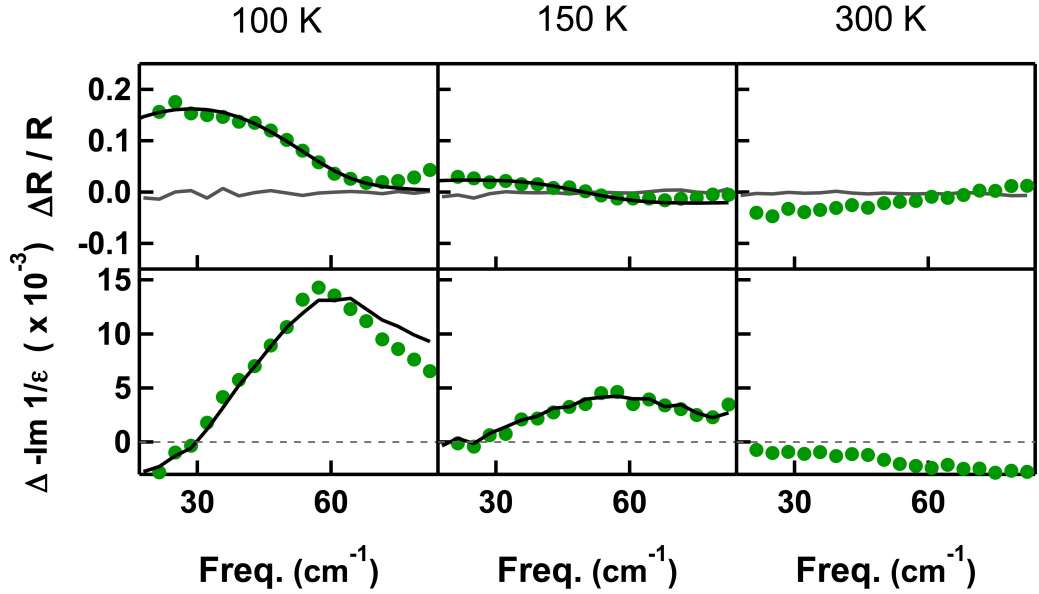

**Fig. S1:** Temperature dependence of the light-induced reflectivity changes $\Delta R/R_0$ in YBa$_2$Cu$_3$O$_{6.6}$ (green dots), measured above $T_C$. As described in the main paper, the sample was excited with 300-fs pulses at 15-µm wavelength, polarized along the c-axis. The data are taken at +0.8 ps time delay. At negative time delay (grey solid lines), the sample does not show any response. The black solid lines are fits to the data, obtained by applying the Bruggeman effective dielectric function $\tilde{\varepsilon}_E(\omega)$ for an inhomogeneous medium [2,3].

## S2: Time-resolved resonant soft x-ray diffraction

The femtosecond resonant soft x-ray diffraction experiment was carried out at the SXR beamline of the LCLS free electron laser (FEL).

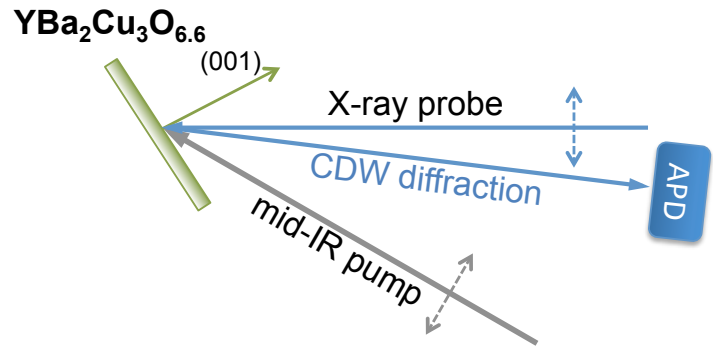

**Fig. S2:** Sketch of the experimental setup used for the femtosecond RXSD experiment at the LCLS free electron laser.

Figure S2 sketches the experimental geometry, combining 400-fs, 4-mJ/cm2 mid-infrared pump pulses at 15-µm wavelength with ~100-fs x-ray probe pulses at 931.5 eV photon energy. The geometry of the x-ray beams is defined by the diffraction condition at the $YBa_2Cu_3O_{6.6}$ in-plane wavevector at $q_{//}$ = 0.315. The FEL provides horizontal polarization, thus lying in the in the scattering plane. For an efficient excitation of the apical oxygen phonon mode, the mid-IR beam is p-polarized and focused non-collinearly onto the sample thus providing the highest possible amount of pump intensity (~80 %) polarized along the sample $c$ axis. The angle of non-collinearity is limited by an increase in the spot size on the sample (reduced overall excitation fluence) and the loss of temporal resolution, though.